\documentclass[twocolumn]{revtex4}
\usepackage{graphicx,float,gensymb}
\usepackage{pifont}
\usepackage{color}

\begin{document}

\title{\huge{Characterization of the irregularity of a terrain using fractal dimension of lakes' boundaries}}
\author{{Nakul N. Karle$^1$ and Kiran M. Kolwankar$^2$} \\
$^1$Department of Physics, University of Mumbai, Vidyanagari, Santacruz(E), Mumbai 400 098\\
$^2$Department of Physics, Ramniranjan Jhunjhunwala College, Ghatkopar(W), Mumbai 400 086\\
E-mail: nakulkarle@gmail.com, kiran.kolwankar@gmail.com
}

\begin{abstract} 

Even though many objects and phenomena of importance in geophysics have been shown to have fractal character, there are still many of them which show self-similar character and yet to be studied.  The objective of the present work is to demonstrate that the fractal dimension of the boundary of a natural water body can be used to shed light on irregularity as well as other properties of a region. Owing to easy availability of satellite images and image processing softwares this turns out to be a handy tool. In this study, we have analyzed several lakes in India mostly around the Western Ghats region. We find that the fractal dimension of their boundaries for the length scales between around 40 meters to 2 kilometers, in general, has broad variation from 1.2 to 1.6. But when they are grouped into three categories, viz., lakes along the ridge of Western Ghats, lakes in the planes and lakes in the mountain region,
we find the first two groups to have a narrower distribution of dimensions.

\end{abstract}

\maketitle

The early measurements of the fractal dimension of the coastline of Britain was used 
by Mandelbrot~\cite{BB} to popularize the concept of fractals. Since then several
objects and phenomena have been studied and shown~\cite{JF,DL,MF,GA} to have fractal nature.
Especially in geophysics~\cite{DL} the fractals and multifractals have been omnipresent.
The coastlines, river networks~\cite{TBR,RR}, porous rocks~\cite{Tho,Sah}, distribution of 
ores~\cite{Car,Tur} being some examples.
However, many aspects still remain to be studied where the fractal nature can turn out to be
a powerful characterizing tool.

Attempts to understand these empirical observations of fractal nature of different geophysical
objects has led to deep theoretical body of work. For instance, the fractal nature of 
the river networks has given rise to the concept of optimal channel networks~\cite{RRIB}.
The modeling of porous rocks involves concepts from the theory of percolation~\cite{Sah}.
In order to understand the formation of fractal coastlines, Sapoval et al.~\cite{BS} argued 
that the coastline evolves so as to balance two competing forces, one by the waves
eroding the interface and other is the damping of waves by the irregular coastline.

In this work, we study the fractal dimension of the boundaries of natural water
bodies. To the best of our knowledge it has not yet been studied~\footnote{When this manuscript was in preparation we came across Ref.~\cite{SPMKB} in which fractal dimension of lakes, and not lakes'
 boundaries, was studied and higher values of dimensions were reported which in fact should be equal to 2.}. 
We demonstrate that it yields a simple characterization of the irregularity
of the surrounding terrain which can be of importance in geophysics. The irregularity
of mountain surface has attracted many, from geologists to computer scientists.
Several mathematical as well as computer models exist which attempt to generate the irregularity of the mountain surface as authentically as possible. Mandelbrot~\cite{BB} has emphasized
studying some aspects of lakes in order to verify the fractional Brownian motion based model of
mountains. Also, results about the coastlines can not be directly used for the lakes' boundaries as the processes involved in dictating the boundary of a lake are different and as argued in 
Ref.~\cite{Bau} their dimensions have no relation in that case. Though the 
erosion is common in two, clearly the impact of waves is not important here whereas
phenomena like landslides and also the history of formation of the particular mountain
could have important role to play. Moreover, in the studies involving the river networks,
the quantities of interest are quite different, for example, the basin boundary, the drainage
area and their relations to network statistics. However, in the case of lakes' boundary the
non-fluvial part of the mountain would also be important which is not the case in the studies involving river networks.

\begin{figure} 
\includegraphics[width=8.5cm]{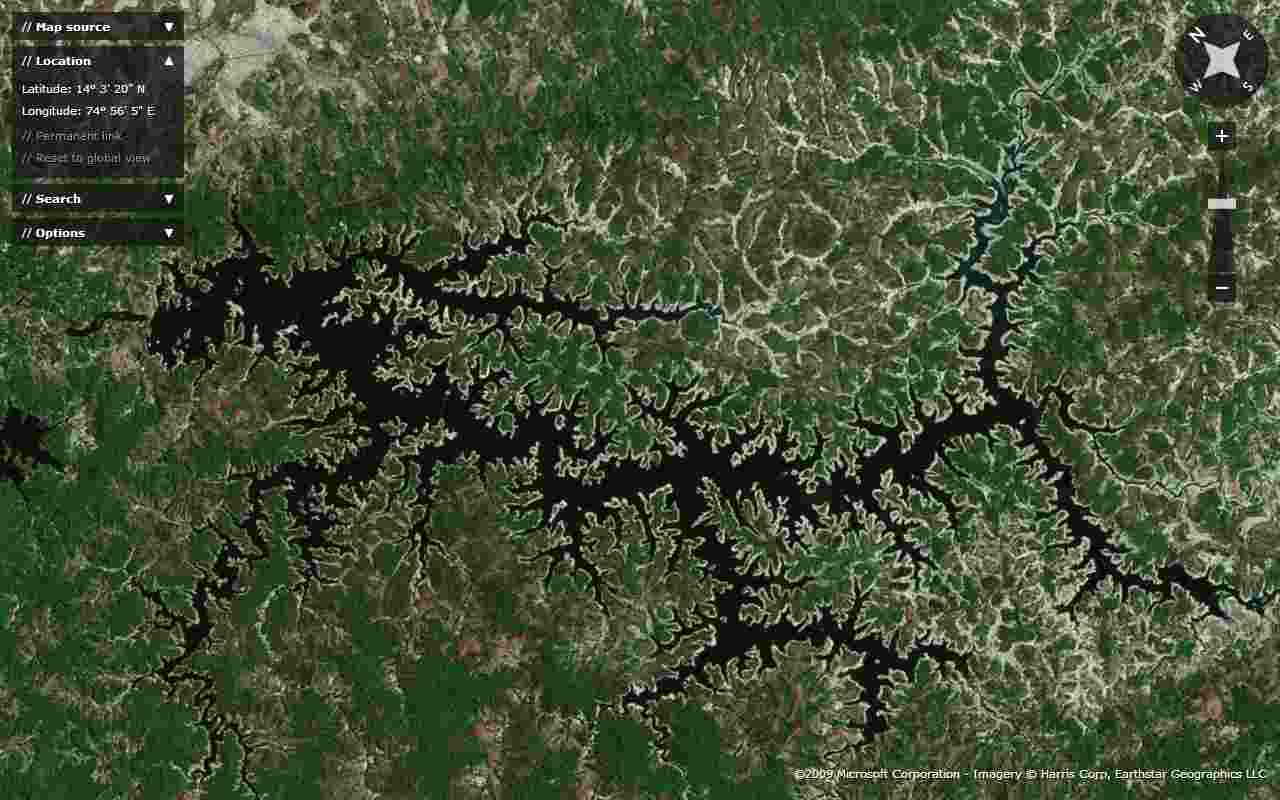}
\caption{Satellite image of the Noanamakki Reservoir, Shimoga, Karnataka, India.}
\label{nr}
\end{figure}

The fractal dimension of river networks has been studied which indirectly gives
a way to quantitatively characterize the irregularity of mountain ranges on a much larger
length scales.
Another way to characterize the irregularity of the mountain surface would be to extract
the outline of the mountains from a photograph and then find its fractal dimension.
While recording a full three dimensional surface topography is technologically
a demanding task, another simple way to obtain the insight is to look at the
fractal dimension of the boundaries of natural water reservoirs. With readily
available satellite images and the image analysis softwares this turns out to be
a handy tool. Moreover, this offers information about the irregularity of the 
horizontal slices of the mountains which is a direction orthogonal to that
offered by the outline. Moreover, this method yields local information
over a range of few kilometers, as against that obtained from the river network,
 thus making it possible to study the variation
from region to region.

\begin{figure} 
\includegraphics[width=8.5cm]{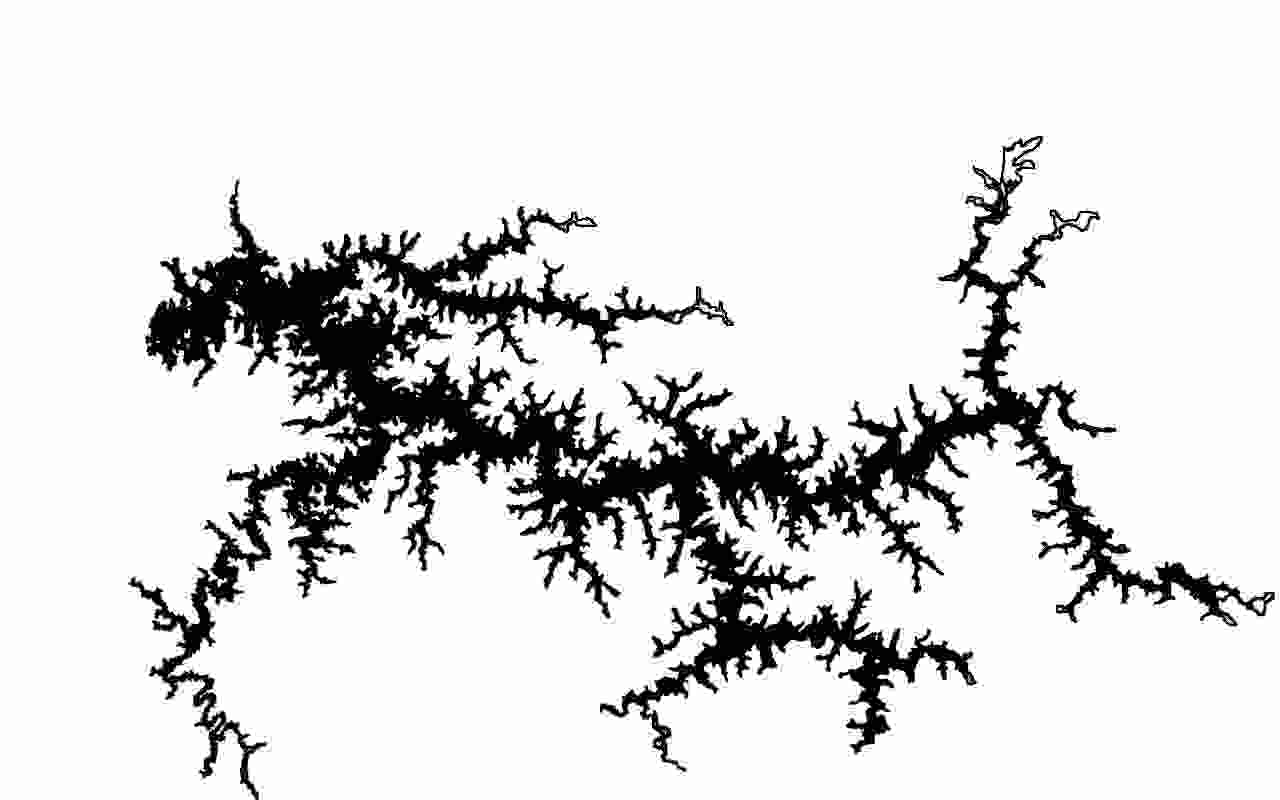}
\caption{The binary image of the reservoir.}
\label{bn}
\end{figure}

\begin{figure} 
\includegraphics[width=8.5cm]{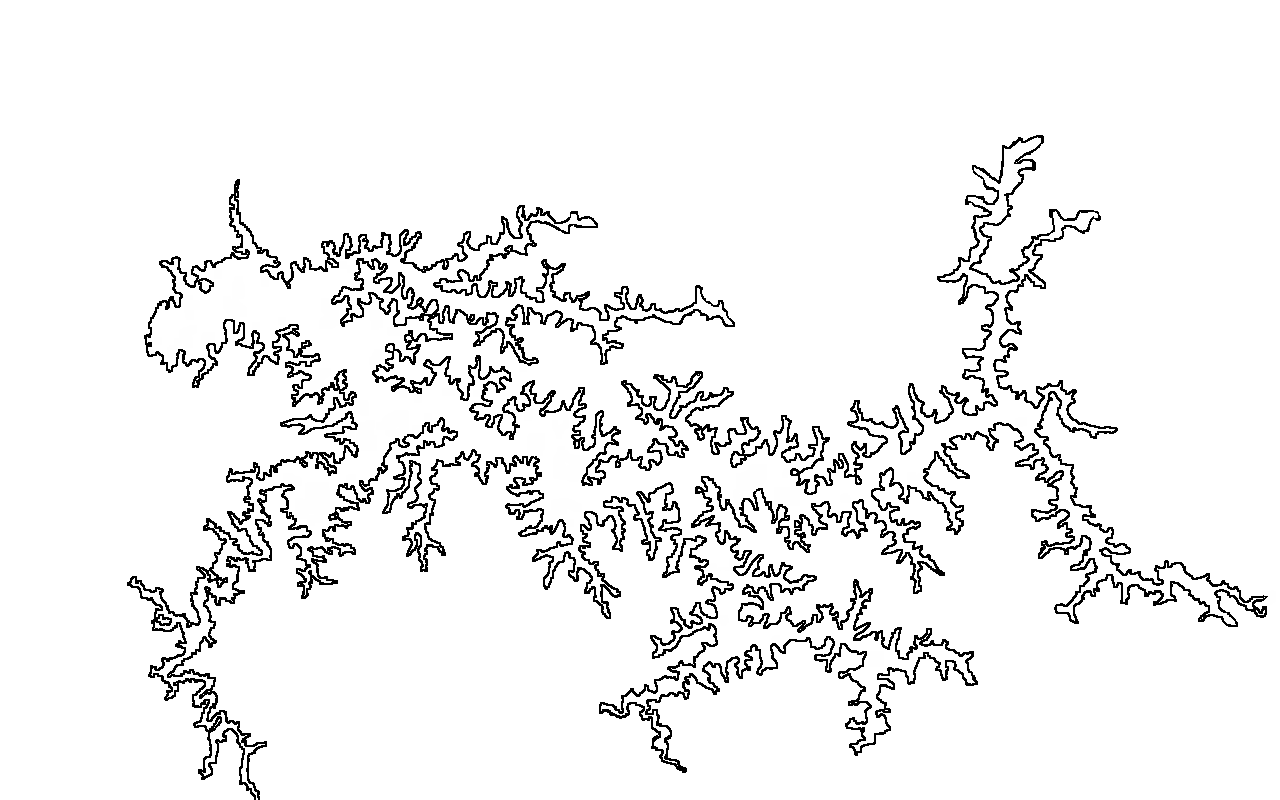}
\caption{The extracted boundary of the binary image.}
\label{ol}
\end{figure}

Now we describe the procedure we followed in order to find the fractal dimension of
a lake's boundary. As an example we take the Noanamakki Reservoir whose image is
shown in Fig.~\ref{nr}.
We first extract the boundary of the lake one is analyzing. In order
to achieve this, the water body is first colored black and then the binary image consisting
only the water body is extracted (Fig.~\ref{bn}). Then the edge detection algorithm is used to obtain the
boundary of the lake as depicted in Fig.~\ref{ol}.

\begin{figure}
\centering
\includegraphics[width=8cm]{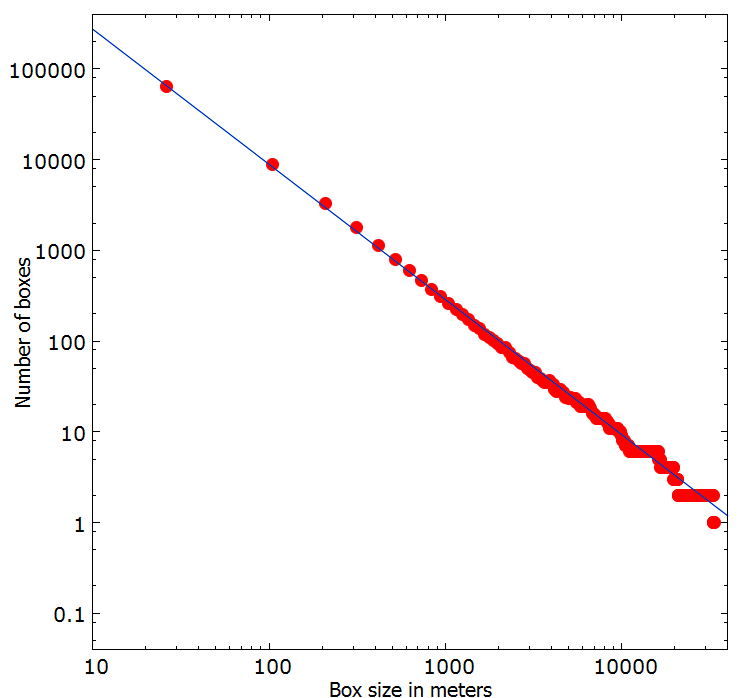}
\caption{Log-log plot of the number of boxes needed to cover the boundary with a given size of the box.
Best line fit with slope 1.60.}
\label{graph}
\end{figure}

The next step is to analyze the boundary again by using readily available fractal analysis tools. We have used ImageJ~\cite{IJ} and Fractalyse~\cite{FS} softwares and we find the boundary to be fractal with a good power law fit from 40 meters to 2000 meters (Fig.~\ref{graph}).

We have analyzed several lakes in India. They were of different sizes, from 8 km to 70 kms.
The lake shown in Fig.~\ref{nr} is around 30 kms wide. The large lakes were divided in parts
and magnified so that the length per pixel was around 15 meters. This allowed us 
to obtain finer details of the boundary. Then the power law exponent from the
pixel size 8 to 128 was found. This is the reported value of the dimension. And
the error estimates were obtained from the change in this exponent when power
law was fitted over a little larger or smaller range of pixels. 

\begin{figure}
\centering
\includegraphics[width=8cm]{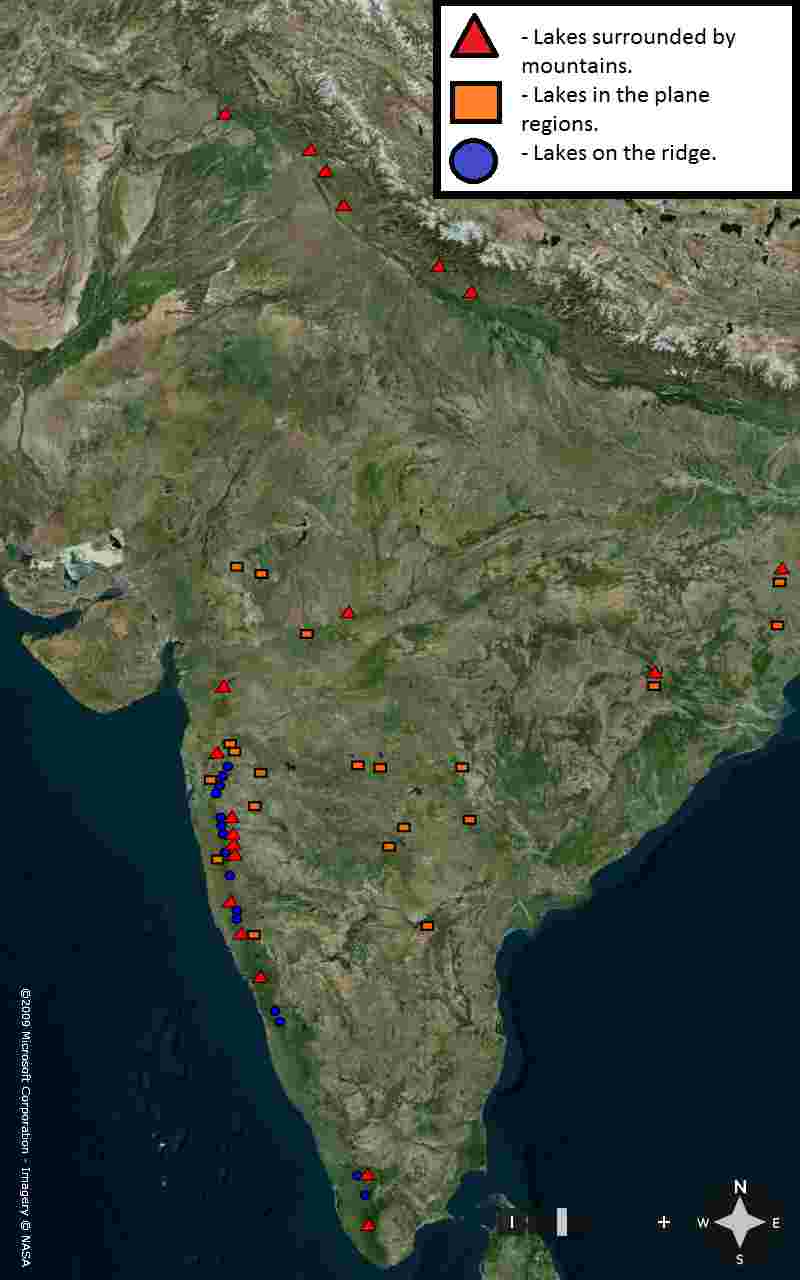}
\caption{Location of the lakes analyzed: Lakes surrounded by mountains (\ding{115}), Lakes in the plane regions (\ding{110}) and Lakes on the ridge (\ding{108}).}
\label{mp}
\end{figure}

We have divided the lakes into three groups. The first 
group consists of the lakes along
the ridge of the Western Ghats (or Sahyadri) mountain range, the second one includes the lakes in the mountains but away from 
the ridge and the third group consists of the ones in the planes. 
We call the ridge of the Sahyadri as that strip which has a width of 10'
from the western edge of the mountain range. Also, the term \emph{lakes in the planes} is only relative
meaning that the mountains around the lake are not very high, usually much less than but not definitely
higher than 100 meters in height.
The lakes were selected randomly with a very few exceptions (less than 10\%).
These were either owing to unclear image or difficulty in classifying the lake.
Also, all the lakes considered in this work happen to be artificial, that is, they were created by damming (they are natural only in the sense that there are no other walls constructed
along the peripheri).
The locations of the lakes analyzed
are depicted on a map in the Fig.~\ref{mp}. The findings are tabulated in three tables below.

We find that, in general, the value of the dimension of the lakes' boundary varies from 1.2 to 1.6.
But when the lakes are grouped as described above a pattern emerges. The lakes along the ridge
and the ones in the plane region have a narrow distribution of dimension. The average value of
the dimension for the lakes along the ridge is 1.4 with a standard deviation of 0.05 and the average
value of the dimension of the lakes in the planes is 1.35 and the standard deviation is 0.06. 
Whereas for the lakes in the mountain region though the average dimension is same as 1.35 there
is a broader distribution with a standard deviation of 0.1.

\begin{table}
\caption{Lakes situated in the plane region.}

\begin{tabular}{|l|c|c|}
\hline
\,\,\,\,\,\,\,\,\,\,\, Name & Latitude and & Fractal \\
& Longitude & Dimension \\
\hline

\,\, Siddeshwar \,\,&\,\, 19{\degree}37'N 76{\degree}53'E \,\,&\,\, 1.40  $\pm$ 0.03 \,\,\\
\hline
\,\, Dhamani \,\,&\,\, 19{\degree}54'N 73{\degree}05'E \,\,&\,\, 1.35 $\pm$ 0.02 \,\,\\
\hline
\,\, Barvi \,\,&\,\, 19{\degree}10'N 73{\degree}21'E \,\,&\,\, 1.37 $\pm$ 0.03 \,\,\\
\hline
\,\, Isapur \,\,&\,\, 19{\degree}47'N 72{\degree}22'E \,\,&\,\, 1.36 $\pm$ 0.03 \,\,\\
\hline
\,\, Jaisamand \,\,&\,\, 24{\degree}15'N 73{\degree}58'E \,\,&\,\, 1.27 $\pm$ 0.02 \,\,\\
\hline
\,\, Kamalpur \,\,&\,\, 17{\degree}49'N 77{\degree}20'E \,\,&\,\, 1.22 $\pm$ 0.01 \,\,\\
\hline
\,\, Nizam Sagar \,\,&\,\, 18{\degree}08'N 78{\degree}00'E \,\,&\,\, 1.29 $\pm$ 0.03 \,\,\\
\hline
\,\, Bagh \,\,&\,\, 21{\degree}01'N 80{\degree}28'E \,\,&\,\, 1.33 $\pm$ 0.02 \,\,\\
\hline
\,\, Mukutmanipur \,\,&\,\, 28{\degree}59'N 86{\degree}46'E \,\,&\,\, 1.40 $\pm$ 0.04 \,\,\\
\hline
\,\, Maithon \,\,&\,\, 23{\degree}48'N 86{\degree}47'E \,\,&\,\, 1.37 $\pm$ 0.03 \,\,\\
\hline
\,\, Som Kamla \,\,&\,\, 23{\degree}58'N 73{\degree}59'E \,\,&\,\, 1.31 $\pm$ 0.03 \,\,\\
\hline
\,\, Mula \,\,&\,\, 19{\degree}18'N 74{\degree}34'E \,\,&\,\, 1.46 $\pm$ 0.02 \,\,\\
\hline
\,\, Darna \,\,&\,\, 19{\degree}45'N 73{\degree}43'E \,\,&\,\, 1.34 $\pm$ 0.01 \,\,\\
\hline
\,\, Yeldari \,\,&\,\, 19{\degree}45'N 76{\degree}42'E \,\,&\,\, 1.41 $\pm$ 0.03 \,\,\\
\hline
\,\, Singur \,\,&\,\, 17{\degree}49'N 77{\degree}54'E \,\,&\,\, 1.32 $\pm$ 0.03 \,\,\\
\hline
\,\, Yashwant Sagar \,\,&\,\, 18{\degree}16'N 74{\degree}57'E \,\,&\,\, 1.30 $\pm$ 0.06 \,\,\\
\hline
\end{tabular}
\end{table}

To conclude, the analysis of the lakes from all over India shows that the boundary has
a self similar structure atleast over the length scales ranging from 40 meters to 2 kilometers.
The values of the fractal dimension varies from lake to lake. We have divided the
lakes into groups: the lakes in the Western Ghats region and lakes in the plane regions. The ones
in the mountain region are further divided into two groups: the ones along the ridge
of the Western Ghats and ones away from the ridge. We find that the average value
and the spread of the fractal dimension varies among these three groups. The lakes in
the planes as well as those along the ridge tend to have a higher values of the dimension
with a small spread in the values. Whereas those lakes in the mountains but away from
the ridge have smaller values with a large deviation.

It is important to understand the reasons leading to these values of
the dimension through physical modeling. In the similar approach to understand the 
irregularity of the coastlines one is led to believe~\cite{BB} that the self-stabilization
between the process of erosion by the waves bombarding the coast and damping of the waves
by the irregularity of the interface leads to a unique value of the dimension. But in the
case of the boundaries of lakes, clearly waves do not have such a vital role to play.

\begin{table}
\caption{Lakes situated on the ridge.}

\begin{tabular}{|l|c|c|}
\hline
\,\,\,\,\,\,\,\,\,\,\, Name & Latitude and & Fractal \\
& Longitude & Dimension \\
\hline

\,\, Varasgaon \,\,&\,\, 18{\degree}22'N  17{\degree}35'E \,\,&\,\, 1.34  $\pm$ 0.01 \,\,\\
\hline
\,\, Bandhardhara \,\,&\,\, 19{\degree}31'N 73{\degree}43'E \,\,&\,\, 1.41 $\pm$ 0.01 \,\,\\
\hline
\,\, Panshet \,\,&\,\, 18{\degree}18'N 19{\degree}38'E \,\,&\,\, 1.38 $\pm$ 0.03 \,\,\\
\hline
\,\, Dajipur \,\,&\,\, 16{\degree}23'N 73{\degree}54'E \,\,&\,\, 1.40 $\pm$ 0.02 \,\,\\
\hline
\,\, Koyna \,\,&\,\, 17{\degree}52'N 73{\degree}44'E \,\,&\,\, 1.43 $\pm$ 0.01 \,\,\\
\hline
\,\, Manikdoh \,\,&\,\, 19{\degree}15'N 73{\degree}45'E \,\,&\,\, 1.39 $\pm$ 0.02 \,\,\\
\hline
\,\, Idukki \,\,&\,\, 09{\degree}48'N 76{\degree}59'E \,\,&\,\, 1.46 $\pm$ 0.02 \,\,\\
\hline
\,\, Bhatghar \,\,&\,\, 18{\degree}11'N 17{\degree}44'E \,\,&\,\, 1.42 $\pm$ 0.02 \,\,\\
\hline
\,\, Chandoli \,\,&\,\, 17{\degree}10'N 73{\degree}47'E \,\,&\,\, 1.33 $\pm$ 0.02 \,\,\\
\hline
\,\, Kalammawadi \,\,&\,\, 16{\degree}18'N 17{\degree}58'E \,\,&\,\, 1.36 $\pm$ 0.01 \,\,\\
\hline
\,\, Mani \,\,&\,\, 13{\degree}40'N 75{\degree}04'E \,\,&\,\, 1.51 $\pm$ 0.02 \,\,\\
\hline
\,\, Upper Bhavani \,\,&\,\, 11{\degree}14'N 76{\degree}42'E \,\,&\,\, 1.38 $\pm$ 0.01 \,\,\\
\hline
\,\, Savehaklu \,\,&\,\, 13{\degree}45'N 75{\degree}00'E \,\,&\,\, 1.43 $\pm$ 0.03 \,\,\\
\hline
\,\, Thokarwadi \,\,&\,\, 18{\degree}54'N 73{\degree}71'E \,\,&\,\, 1.45 $\pm$ 0.02 \,\,\\
\hline
\end{tabular}
\end{table}

In mountains, the irregularity of the boundary is clearly dictated by the irregularity 
of the mountain itself. Which in turn is decided by the process of the formation of
the mountain range. The erosion processes, landslides, weathering, fracture patterns and also the movement of tectonic 
plates could have a role in this
other than reasons specific to that mountain range. 
The difference between the lakes along the ridge and elsewhere in the mountains could
be attributed to the fact that the ridge of the Western Ghats is the origin of the volcanic eruptions. 
It would be interesting to construct
a physical model of mountain formation which will reproduce the values of the
fractal dimension measured here. 
What is surprising is the fact that even in the planes the values of dimensions
are spread over a small range. Clearly, here the properties of the soil will have a 
major role to play in deciding the dimension. 

The results of the work by Morais et al.~\cite{Mor} on the erosion in the correlated
landscapes could be of use in explaining some of the observations here. In~\cite{Mor},
the authors find that the dimension of the interface depends on the spatial
correlations in the lithography parameter of the landscape leading to various values
of the dimension. This could offer an explanation for the spread of values of the
dimension we find.

\begin{table}
\caption{Lakes surrounded by the mountains.}

\begin{tabular}{|l|c|c|}
\hline
\,\,\,\,\,\,\,\,\,\,\, Name & Latitude and & Fractal \\
& Longitude & Dimension \\
\hline

\,\, Tilari \,\,&\,\, 15{\degree}44'N  74{\degree}06'E \,\,&\,\, 1.30  $\pm$ 0.03 \,\,\\
\hline
\,\, Thenmala \,\,&\,\, 08{\degree}55'N 77{\degree}07'E \,\,&\,\, 1.41 $\pm$ 0.01 \,\,\\
\hline
\,\, Ranjit Singh \,\,&\,\, 32{\degree}29'N 75{\degree}46'E \,\,&\,\, 1.40 $\pm$ 0.01 \,\,\\
\hline
\,\, Ramganga \,\,&\,\, 29{\degree}35'N 78{\degree}45'E \,\,&\,\, 1.39 $\pm$ 0.01 \,\,\\
\hline
\,\, Tawa \,\,&\,\, 23{\degree}31'N 78{\degree}02'E \,\,&\,\, 1.44 $\pm$ 0.01 \,\,\\
\hline
\,\, GobindSagar \,\,&\,\, 31{\degree}24'N 76{\degree}29'E \,\,&\,\, 1.33 $\pm$ 0.01 \,\,\\
\hline
\,\, Hirakud \,\,&\,\, 21{\degree}37'N 83{\degree}47'E \,\,&\,\, 1.36 $\pm$ 0.02 \,\,\\
\hline
\,\, Urmodi \,\,&\,\, 17{\degree}40'N 73{\degree}52'E \,\,&\,\, 1.23 $\pm$ 0.03 \,\,\\
\hline
\,\, Noanamakki \,\,&\,\, 14{\degree}03'N 74{\degree}56'E \,\,&\,\, 1.60 $\pm$ 0.01 \,\,\\
\hline
\,\, Kanher \,\,&\,\, 17{\degree}46'N 73{\degree}51'E \,\,&\,\, 1.25 $\pm$ 0.01 \,\,\\
\hline
\,\, Mangla \,\,&\,\, 33{\degree}10'N 73{\degree}47'E \,\,&\,\, 1.44 $\pm$ 0.01 \,\,\\
\hline
\,\, Mullaperiyar \,\,&\,\, 09{\degree}32'N 07{\degree}11'E \,\,&\,\, 1.48 $\pm$ 0.02 \,\,\\
\hline
\,\, Ukai \,\,&\,\, 21{\degree}17'N 73{\degree}43'E \,\,&\,\, 1.35 $\pm$ 0.02 \,\,\\
\hline
\,\, Upper Vaitarna \,\,&\,\, 19{\degree}50'N 73{\degree}32'E \,\,&\,\, 1.41 $\pm$ 0.04 \,\,\\
\hline
\,\, Tehri \,\,&\,\, 30{\degree}25'N 78{\degree}27'E \,\,&\,\, 1.21 $\pm$ 0.04 \,\,\\
\hline
\end{tabular}
\end{table}

Clearly, in the case of reservoirs located in the mountains the irregularity in its boundary is a consequence of that of the mountains. Which, in turn, depends on the geological details of formation of that mountain. The boundary of a lake can be looked upon as a 
horizontal cross section of the mountain. Thus its dimension is the dimension of the cross section. Now if we find that the dimension of such cross sections do not depend much on where the cross section is taken, that is the dimension of the boundary does not depend on the height, then the dimension of the lake’s boundary gives us an estimate of the dimension of the mountain~\cite{KK}.
By general theory the dimension of a typical cross section of a fractal surface can
be expected to have dimension one less than that of the surface~\cite{KF}. Thus if $\alpha$ is the dimension of the boundary of a lake then 1+$\alpha$ is the dimension of the mountain surface.

Any good model of mountain formation would have to reproduce this range of values. More analysis would be needed to see if there is any correlation between the height of the lake from the sea level and its fractal dimension or if there is any systematic dependence of the fractal dimension on the region. In future, it is planned to carry out an extensive analysis of lakes all over the globe and
address these and related questions.

\emph{Acknowledgments:} The satellite images analyzed were taken from the Google Earth and Microsoft Bing websites. We thank Professor
Shrinivas Viladkar for discussions and carefully reading the manuscript.

\end{document}